\documentclass[conference]{IEEEtran}
\usepackage{xcolor}
\usepackage[natbib=true, style=ieee, url = true, sorting=none]{biblatex}
\usepackage{amsmath,amssymb,amsfonts}
\usepackage{algorithmic}
\usepackage{graphicx}
\usepackage{textcomp}
\usepackage{svg}
\usepackage{booktabs}
\usepackage{array}

\makeatletter 
\def\blfootnote{\xdef\@thefnmark{}\@footnotetext} 
\makeatother

\def\BibTeX{{\rm B\kern-.05em{\sc i\kern-.025em b}\kern-.08em
    T\kern-.1667em\lower.7ex\hbox{E}\kern-.125emX}}

\begin{document}

\title{Procedural Generation and Games \\at the Dawn of Fault Tolerant Quantum Computing}

\thanks{Vision paper}

\author{\IEEEauthorblockN{1\textsuperscript{st} Daniel Bultrini}
\IEEEauthorblockA{\textit{Moth} \\
Schorenweg 44b, 4144 Arlesheim, Switzerland \\
daniel@mothquantum.com}
\thanks{Vision paper}

\and
\IEEEauthorblockN{2\textsuperscript{nd} James  Wootton}
\IEEEauthorblockA{\textit{Moth} \\
Schorenweg 44b, 4144 Arlesheim, Switzerland \\
james@mothquantum.com}}

\maketitle
\thanks{Vision paper}

\begin{abstract}
Quantum computers have long been more of a toy for researchers than a tool for solving complex problems. However, recent advances in the field make exploiting the advantages of fault-tolerant quantum computers feasible in the next 5 to 10 years. It is now time to begin imagining how such devices could be used in practice for game development and deployment. In this work we identify procedural content generation as a very promising area of application and exploration. We examine a selection of algorithmic approaches used in classical procedural content generation and propose promising quantum algorithms that could provide an alternative approach or a computational advantage. We then end with a hypothetical game that exploits a recent quantum algorithm for computing the Jones polynomial exponentially faster than classical computers could. 
\end{abstract}


\section{Introduction}
Quantum computing is a different approach to computation which promises advantages to solving various, generally complicated, problems. Like any new technology, it is impossible to avoid thinking about how it could be used to create experiences that couldn't be made otherwise. Currently, quantum computers (QCs) are noisy machines that cannot reliably outperform digital computers, but in the coming decade this might change, with many manufacturers promising the first generation of fault-tolerant QCs (FTQC) \cite{aasenRoadmapFaultTolerant2025}. What could these first functional QCs bring to games? We believe that due to the limited availability of early QCs, the most promising applications will be in the area of procedural content generation (PCG), which we will focus most of our discussion on, but it could also bring games that are currently computationally intractable into existance.

This vision paper aims to explore what the potential advent of this new computational technology could mean for PCG and games with a practical lens. This will be done by classifying a large selection of common PCG algorithms, suggesting quantum algorithms (QA) that could supplement or replace them in future quantum-powered methods, determining the level of fault-tolerance required for their utility, and hypothesizing about uncharted territory. But why would we want to use QCs here?

Procedural content generation (PCG) is, in a way, the artistic application of mathematics. As with many art forms, PCG is defined by the strengths and limitations of its medium. 
For PCG, the medium is that of computation, and hence the constraints can be expressed in terms of computational complexity. 
PCG is used throughout the creative industries for many reasons. In the early days of computer gaming it was often used to increase replayability as in \textit{Rogue} \cite{toyRogue1980} but also as a way to create immense worlds that would never fit into the limited storage of early home computers, such as in \textit{Daggerfall} \cite{boumaElderScrollsII1996}. Nowadays the massive size and intricacy of game worlds and the breadth of desirable player-game interaction make PCG an important tool to this day. It can be used in almost every aspect of the creation process, from more abstract parts such as rules and narratives to more traditional uses such as maps, levels and assets - yielding an augmentation to the human creative process and the potential to reduce development cost \cite{withingtonCompressingComparingGenerative2022}. 

To achieve these feats, PCG uses myriad different techniques, from generating simple noise functions to tackling complex constraint satisfaction problems. The prospect of a new paradigm of computing, in which the resources required to solve certain problems are fundamentally changed, could therefore allow a range of new forms of expression within the procedural generation of content.
\blfootnote{Abbreviations: 
CA: Cellular Automata,
CC: Classical Computation,
CSP: Constraint Satisfaction Problems,
DQI: Decoded Quantum Interferometry,
EA: Evolutionary Algorithms,
FTQC: Fault-Tolerant Quantum Computers,
JP: Jones Polynomial,
PCG: Procedural Content Generation,
QA: Quantum Algorithms,
QAOA: Quantum Approximate Optimization Algorithm,
QC: Quantum Computer,
WFC: Wavefunction Collapse}
This is what is potentially offered by quantum computation. The development of QCs was originally motivated by the observation that quantum systems are impractically inefficient to simulate on conventional digital hardware. This includes the broad spectrum of digital hardware, from CPUs to GPUs and from microcontrollers to supercomputers. All are simply referred to as `classical computation' (CC) in the quantum context. Over decades of QA research, it has been proven that QCs -- highly isolated abstract quantum systems that can be precisely controlled -- can do more than just simulate quantum physics. For one thing, they can also efficiently emulate any classical algorithm, meaning that it is possible in principle to run DOOM on quantum hardware~\cite{mortimerQuandoomDOOMQuantum2024}. Not only this, they have been shown to offer improvements in computational complexity for a broad class of problems, from unstructured search to factoring and beyond \cite{nielsenQuantumComputationQuantum2010}. This computational advantage is what is called \emph{quantum advantage}.

This paper will not require an understanding of quantum computing, but will rather focus on the potential advantages and considerations required for the use of QAs in PCG. As such we have chosen to forego an introduction to quantum computing, but we recommend this very approachable introduction \cite{jQuantumAlgorithmImplementations2022}. 

\subsection{Games and PCG on Quantum Computers}

Using QCs for games or making games to explore QCs is not new, although most examples stem almost exclusively from QC game jams and QC hackathons. The definition of what is a quantum game is not set in stone, and there is a difference between a game that uses or explores some aspect of quantum physics \cite{piispanenDefiningQuantumGames2025} and those games that fundamentally require a QC to compute some essential gameplay mechanic or feature. Nonetheless, it is something that is being explored and there is an active community of game makers and researchers working in this field. 

This leads onto the focus of this paper, PCG, which is less explored than games in general in the realm of QC. The interesting thing about PCG is that the algorithms behind many systems can be extremely computationally complex, and might actually benefit from a fault-tolerant QC. To this end, some initial groundwork has been laid \cite{woottonProceduralGenerationUsing2020}. A version of quantum map generation \cite{woottonQuantumProcedureMap2020} exists which limits itself to colouring a territory based on a set of rules and the connectivity of the device determining the `borders' of the territories. The rules are then quantum gates applied between cities and the outcome of computation determines if a territory has grown or been taken over. The final product of this is a coloured map that could be used as the basis for the rest of a game or other mapping to a gameplay mechanic. 

An interesting case is the \emph{wavefunction collapse method} (WFC), an algorithm that borrows its name from the quantum phenomenon. Based on Paul Merrel's model synthesis algorithm \cite{merrellModelSynthesisGeneral2011}, WFC has had two quantum versions proposed \cite{zhangWaveFunctionCollapse2024,heeseQuantumWaveFunction2024}. WFC has been used in a plethora of different tasks, including generating textures and terrain as well as more functional PCG tasks like the layouts of roads, transport networks and puzzles. The interest in recasting the algorithm back to its namesake is likely due to the fact that it is fundamentally a constraint satisfaction solver \cite{karthWaveFunctionCollapseConstraintSolving2017}, and as such it is possible that it could be extended to incorporate large numbers of non-local constraints when cast as a QA, but this can also be done classically.

\section{Fault Tolerant Quantum Algorithms for PCG}

Fault-tolerant QAs can be further subdivided into two categories useful to this discussion, those that take physical constraints of the hardware into account, which we call algorithms for early-FTQCs and those that are meant for completely fault-tolerant machines. The difference is that early-FTQC algorithms will need to be tailored to the exact constraints of a specific early-FTQC, like writing a program for an early computer. As the hardware matures, more abstract mathematical formulations can be used at scale effectively, much like today's computing experience. 

It is expected that  early-FTQCs will be too slow for real-time generation, and so would likely be more useful in workloads where they are used for offline generation. It is not impossible that algorithms could be run during gameplay to build a buffer of generated content, but we believe that real-time use is outside the scope of a reasonable implementation of early-FTQCs, and potentially FTQCs \cite{proctorBenchmarkingQuantumComputers2025} in most contexts. Of course, some deterministic or probabilistic algorithms that require one circuit and few queries do exist, such as factoring \cite{nielsenQuantumComputationQuantum2010} and finding the Jones polynomial \cite{laakkonenLessQuantumMore2025} could be used in real-time. 

\subsection{The cans and can'ts of quantum computing}
PCG has several hurdles that algorithms must overcome. A famous one is the \emph{thousand bowls of oatmeal} problem \cite{rabiiWhyOatmealCheap2023}, where different samples of generated content differ only in ways that are meaningless to the player, rather than feeling like purposefully different experiences. Though this is a meta-problem that requires many algorithms and careful construction to truly overcome, with more command over navigating the possibility space we can hope to add meaningful differences to our oatmeal. QCs could offer this by making it feasible to search a larger space, but the fact remains that the search space itself must have interesting solutions. This also brings another problem to light, that of the construction of the search space \cite{togeliusProceduralContentGeneration2013}, which underlies many PCG algorithms.  Other considerations include multi-level PCG (both in scale and in sequence), controllability of the algorithm, and methods to verify whether outputs are of high quality. 

The problems described above stem from the fact that quantifying concepts like `fun, `beauty' and `challenge' is difficult. Of course, there are methods that have been developed especially with the advent of machine learning. Regardless, QCs alone are not expected to be a silver bullet - they may greatly speed up routines or subroutines. This could allow for the exploration of much larger and more complex search spaces without suffering from the \emph{curse of dimensionality} as much as CCs do, but finding solutions that are varied and interesting requires that we can translate those concepts. So when looking for where and how QCs can help, it will always be important to keep in mind that the knowledge gained in classical algorithms and theory will always be relevant. 

\subsection{What constitutes a quantum advantage?}
If one algorithm can solve a problem with input size $N$ faster than another, and do so such that this speedup increases with increasing $N$, we say that the former algorithm has an \emph{advantage}. If a QA outperforms the best possible classical one for the same problem, we therefore talk of a \emph{quantum advantage}. The most naive example is of a QC with a random set of $O(N)$ layers of universal quantum gates between $N$ qubits, which requires $O(N)$ space and time complexity on a QC but which is exponentially more difficult to simulate classically as $N$ increases. However, similarly random circuits composed of non-universal gates, or non-random circuits composed of universal gates, can sometimes have structure that allows efficient classical simulation.

As such, a QA with an advantage in PCG must fulfill the following criteria: it must be hard to simulate, and it must solve a useful problem. 

To have an idea of what a minimal quantum advantage is, one must first begin by mentioning that there is always the possibility of implementing \textit{Grover's Algorithm} \cite{nielsenQuantumComputationQuantum2010}, which finds a solution to any programmable set of conditions over a search space of $N$ entries, returns an answer in $O(\sqrt{N})$ time. Attempting the same with a classical search would take $O(N)$ time to find the entry that satisfies the conditions of the problem, giving Grover's algorithm a square-root quantum advantage in this most general case. This assumes that both approaches use the same efficient classical algorithm to check the conditions of the problem. It follows that for problems in which a QA can also provide an advantage in this checking step, the quantum advantage will increase, but so far no examples of this exist. These considerations give us a baseline gain in efficiency for any search problem. However, the overheads of fault-tolerance may swallow this advantage for problems of any practical size. It should therefore primarily serve as a motivation to look for more substantial speedups in specific cases. Specific values of $N$ at which a quadratic advantage might be evident is hard to estimate and might even disappear in more intricate pipelines~\cite{agliardiConditionsQuadraticQuantum2025}. 

This is due to the issue of the \emph{end-to-end} complexity. Even if elements within QAs have exponential advantage, inefficient data encoding and decoding of the answer can cause the advantage to be lost completely. Such analysis is not straightforward, and requires rigorous analysis of how data loading algorithms and processing algorithms combine to either maintain or overwhelm the advantage when applied to real-world problems~\cite{agliardiConditionsQuadraticQuantum2025}. As such, any algorithm that is proposed must eventually be analyzed in the context of the full pipeline of the PCG task. This is not something that can be done in this paper, but something similar has been done in a more abstract context \cite{dalzellQuantumAlgorithmsSurvey2023}. In general, in order of most expensive to least expensive quantum resources we have:
\begin{enumerate}
    \item Number of qubits. 
    \item Depth of quantum circuits. 
    \item Number of samples. 
    \item Classical pre- and post-processing.
\end{enumerate} 

Although this paper will focus mostly on fundamental algorithms, it is undeniable that various machine learning models have been used to great effect in PCG \cite{guzdialProceduralContentGeneration2022}. There is a large body of work on quantum machine learning with analogues to most techniques used on the classical side, but it remains an open question whether there is an advantage to using this over classical machine learning, especially on classical data \cite{bowles2024better}. However, as it may be employed to solve more complex underlying problems it could prove useful, so it is worth having this approach in mind. This is because some data is more efficiently encoded in a quantum system. In addition to quantum data, there is evidence that grammar-based information such as natural language has an efficient quantum encoding \cite{karamlouQuantumNaturalLanguage2022}.

Table \ref{tab:pcg_algorithms_fullwidth} summarizes a selection of key classical PCG problems and the related QAs that will be discussed in depth in the rest of this work.

\begin{table*}[h] 
    \centering 
    \caption{Summary of Classical and Quantum Algorithms for PCG Problems and Their Speedups}
    \label{tab:pcg_algorithms_fullwidth}
    \begin{tabular}{>{\raggedright\arraybackslash}p{0.18\textwidth} >{\raggedright\arraybackslash}p{0.18\textwidth} >{\raggedright\arraybackslash}p{0.18\textwidth} >{\raggedright\arraybackslash}p{0.38\textwidth}}
    \toprule
    \textbf{Problem} & \textbf{Exemplary Classical Algorithm} & \textbf{Exemplary Quantum Algorithm} & \textbf{Key takeaways} \\
    \midrule
    Generic Search within a Solution Space & 
    General Search (e.g., iterating through N entries) & 
    Grover's Algorithm \cite{nielsenQuantumComputationQuantum2010} & 
    Quadratic (O(N) to O($\sqrt{N}$) queries), baseline quantum advantage, likely insufficient for most use cases. \\
    \midrule
    Exact Constraint Satisfaction & 
    Backtracking, Answer Set Programming \cite{smithAnswerSetProgramming2011}, various heuristics \cite{bengioMachineLearningCombinatorial2021} & 
    Quantum Backtracking \cite{martielPracticalImplementationQuantum2020}, Quantum Approximate Optimization Algorithm (QAOA) \cite{dalzellQuantumAlgorithmsSurvey2023} and DQI \cite{jordanOptimizationDecodedQuantum2024} & 
    Quantum backtracking has a quadratic speedup, but can exploit the graph structure of a problem unlike Grover's algorithm ($O(\sqrt{T})$ instead of $O(T)$).
    QAOA does not have a proven advantage, and the literature is conflicting on whether this will materialize. For the specific case of sparse max-XORSAT, the recently discovered DQI algorithm was found to outperform generalized classical methods, but not specialized ones. However, DQI's exponential advantage for the polynomial intersection problem shows that there is potential for speedups. \\
    \midrule
    Classical Stochastic Search & 
    Genetic/Evolutionary Algorithms \cite{togeliusSearchbasedProceduralContent2011} & 
    Quantum Genetic/Evolutionary Algorithms \cite{ibarrondoQuantumVsClassical2022}& 
    Advantage shown in some formulations for small problem sizes; unclear for larger problems. Fundamentally different from classical versions, so can be thought of as an alternative approach. \\
    \midrule
    Grammars and Automata & 
    Classical Grammars (L-systems, graph grammars) and Automata (cellular automata) \cite{shakerProceduralContentGeneration2016} & 
    Quantum Grammars and Automata \cite{mooreQuantumAutomataQuantum2000} & 
    Rather than a direct speedup for the same tasks, but instead offers a different generative paradigm that enhances the set of tools and approaches. May display a
    potential speedup in specific sub-steps (e.g., state amplification of populations); primarily yields different patterns or processes. \\
    \midrule
    Various PCG Tasks using Machine Learning & 
    Classical Machine Learning & 
    Quantum Machine Learning \cite{wangComprehensiveReviewQuantum2024} & 
    An open question regarding advantage, especially with classical data. Potential for better performance on quantum data (e.g. chemical wavefunctions), but classical PCG might not be representable in such a form. Both quantum and classical machine learning are not discussed in detail this work. \\
    \bottomrule
    \end{tabular}
    \end{table*}

\subsection{Constraint Satisfaction}

Constraint Satisfaction Problems (CSPs) are fundamental to many PCG problems. They provide a formal framework for defining the desired properties of generated artefacts, such as ensuring coherent transitions in terrain generation or guaranteeing a solvable path to a goal in level design. Constraints are typically expressed as a collection of logical statements that any artefact produced by a PCG algorithm must satisfy. Traditionally, techniques like \emph{answer set programming} have been employed to formalize and solve such problems \cite{shakerProceduralContentGeneration2016}. Once the conditions are established, any proposed solution must meet all specified criteria (a satisfiability problem, SAT) or, in more flexible scenarios, a maximal subset of them (a maximum satisfiability problem, Max-SAT). The core challenge then lies in devising an algorithm to generate candidate solutions that adhere to these constraints.

\subsubsection{Classical Approaches to CSPs}

The only completely general and potentially exact technique for solving CSPs is through search algorithms. For the purpose of this discussion, classical search algorithms can be broadly categorized into \emph{exact} and \emph{stochastic} methods. The search space for a given problem is typically characterized by the properties desired in the final solution, which can be further quantified by an objective or fitness function in the case where multiple solutions exist.
\begin{itemize}
    \item \textbf{Exact search} methods aim to find an optimal solution by systematically exploring the decision space. This can involve an exhaustive search of every potential point, although for certain problem structures, non-exhaustive exact algorithms exist.
    \item \textbf{Stochastic search} algorithms attempt to bypass the computational cost of exhaustive search through intelligent sampling or evolutionary processes within the search space. Prominent examples include genetic and evolutionary algorithms \cite{togeliusSearchbasedProceduralContent2011}.
\end{itemize}

The baseline exact search method employed in PCG is \emph{backtracking}. Given a set of desirable conditions and construction limits, a tree graph of potential solutions can be constructed, either fully or iteratively. As a backtracking algorithm traverses this graph, it explores potential branches. If a branch leads to a valid partial solution, exploration continues; if it violates a constraint (an invalid branch), the algorithm backtracks to a previous decision point and explores an alternative path. This process is repeated until a desired number of valid solutions are found. The classical complexity of backtracking is heavily influenced by the branching factor of the search tree (i.e., the number of choices available at each decision point). For challenging problems like the $N$-queens problem, the complexity can scale factorially, e.g., $O(N!)$. However, if each decision significantly prunes the remaining search space, exhaustive searches like backtracking can be efficient for finding a single solution.

\subsubsection{Quantum Approaches to CSPs}
 Several QAs that attempt to do better than Grover's algorithm have been developed,  which is known to be able to search the space of solutions in $O(\sqrt{T})$ time, where $T$ is the number of candidate solutions, which usually grows exponentially with the problem size. Some promising QAs that could be applied to CSPs are:
\begin{itemize}
    \item \textbf{Quantum Backtracking} demonstrates efficiency in qubit utilization, capable of representing $2^Q$ paths with $Q$ qubits. However, if a classical backtracking algorithm requires $T$ steps to find a solution, the quantum version generally achieves only a quadratic speedup in the number of queries, specifically $O(\sqrt{T})$ \cite{martielPracticalImplementationQuantum2020}. This speedup is akin to that offered by Grover's algorithm, with the notable distinction that the inherent graph structure of the problem can be exploited. Likely only a FTQC algorithm due to the reliance on oracles. 
    \item \textbf{Quantum Approximate Optimization Algorithm (QAOA)} \cite{dalzellQuantumAlgorithmsSurvey2023} is designed to find approximate solutions to combinatorial optimization problems that can be mapped to quadratic unconstrained binary optimization problems. However, there is currently no definitive proof of a general quantum advantage for QAOA over classical heuristics. Furthermore, its reliance on classical optimization of its quantum circuit may pose practical challenges for real-world PCG applications due to it inhereting all the issues that exist in classical optimization. It is a good candidate for current noisy QCs, but may not be as feasible in early-FTQCs. 
    \item \textbf{Decoded Quantum Interferometry (DQI)} has demonstrated a proven exponential advantage for the specific problem of polynomial intersection \cite{jordanOptimizationDecodedQuantum2024}. Crucially, heuristics suggest the potential to extend DQI's quantum advantage to a broader range of problem classes. This makes DQI a particularly promising candidate for developing new quantum-enhanced PCG methodologies.
\end{itemize}
Looking ahead, it is plausible that future QAs for PCG, much like many contemporary classical algorithms, will rely significantly on problem-specific heuristics.

\subsubsection{Classical Stochastic Search} 
\emph{Evolutionary algorithms} (EAs) and their prominent subset, \emph{genetic algorithms} are commonly used.  In their classical formulation, these algorithms operate by maintaining a population of individuals, each representing a potential solution. The runtime complexity of these algorithms is generally proportional to $g \cdot n \cdot m$, where $g$ denotes the number of generations (iterations), $n$ is the number of individuals within each generation, and $m$ represents the complexity associated with evaluating or manipulating a single individual. A critical component of EAs is the fitness function, which is evaluated for every individual in each generation. The computational cost of this fitness evaluation can be arbitrarily complex, depending on the specifics of the problem. This variability in fitness function complexity makes it challenging to establish a universally quantifiable performance advantage for classical EAs across all PCG applications.

\subsubsection{Quantum Stochastic search}

The direct translation of EA mechanisms to QAs is more nuanced. Since the replication (cloning) of highly fit individuals or the outright discarding of the least fit individuals are not directly reproducible due to the no-cloning theorem \cite{nielsenQuantumComputationQuantum2010}. This theorem and its corollaries forbid the cloning or discarding of unknown quantum states. Consequently, to develop \emph{quantum evolutionary algorithms}, the problem formulation and algorithmic operators must be adapted. For instance, instead of perfect cloning, one might employ imperfect cloning techniques achievable with quantum cloning machines.

As such, quantum evolutionary algorithms are probably not directly comparable to their classical counterparts. Research has indicated that certain formulations of quantum EAs can exhibit an advantage over classical EAs for small-scale problems \cite{ibarrondoQuantumVsClassical2022}. It remains an open question whether this advantage consistently scales to larger, more complex problem instances.

\subsection{Constructive techniques}

Constructive techniques represent a paradigm in Procedural Content Generation (PCG) where solutions are built incrementally, step-by-step. A key characteristic of these methods is that each step typically leads to a valid or partially valid output, which can contribute to their computational efficiency. The advent of quantum computing offers the potential to expand the repertoire of tools and approaches within this paradigm. While a QA employed in a constructive process might itself be classically intractable to simulate, its role could be less about solving a computationally hard problem in the traditional sense and more about facilitating a generative task through novel quantum mechanical means.

\textbf{Grammar-based techniques}, such as L-systems and graph grammars, are powerful and versatile tools in PCG, capable of generating a wide array of content, from botanical structures like plants to more intricate systems of interaction \cite{shakerProceduralContentGeneration2016}. Grammars consist of a set of \emph{production rules} that dictate how \emph{symbols} are transformed. 

\subsubsection{Classical grammars} one begins with an initial string of symbols, known as an \emph{axiom}. For a minimal illustration, consider symbols $A$, $B$, and $C$ with the rules: 1) $A \rightarrow AB$, 2) $B \rightarrow BC$, and 3) $C \rightarrow A$. Starting with the axiom $A$, applying rule 1 yields $AB$; subsequently applying rule 2 to $B$ could yield $ABC$, and so forth. Grammars must be carefully designed, and the symbols must be mapped to tangible objects, properties, or processes whose combinations make sense according to the rules. The design of such grammars can be a complex, time-consuming endeavor, leading to research in methods for automatically generating or evolving grammars \cite{merrellExampleBasedProceduralModeling2023}.

For automatic grammars, such as Merrell's automatic graph grammar \cite{merrellExampleBasedProceduralModeling2023}, a hierarchy of primitive features from an input must be constructed. This is then used to synthesize new, locally similar patterns. While promising, this method can encounter limitations, such as the graph hierarchy becoming excessively large, notably when extending to 3D structures. Although specific QAs for this exact problem have not been researched, the underlying task involves computing graph properties. Many QAs exist for computing various graph properties \cite{leeQuantumAlgorithmsGraph2021}, with several offering non-trivial advantages. While the existence of such an algorithm to the specific problem in Example-Based Procedural Modeling is speculative, the existence of quantum advantages for many graph problems might offer a promising avenue towards this. 

\subsubsection{Quantum Grammars} Beyond enhancing classical grammars, \emph{quantum grammars} can be thought of as a related, but different form of process \cite{mooreQuantumAutomataQuantum2000}. Analogous to classical grammars, quantum grammars operate on symbols and rules. They might begin with an initial quantum statevector as an axiom. The symbols in a quantum grammar could be represented by unitary matrices (e.g., $U_A, U_B, \dots$), and the production rules dictate how these unitary operations are sequenced and applied to evolve the statevector. This application can be controlled quantumly (e.g., using control qubits to determine the sequence of unitaries) or guided by classical choices. After the statevector has evolved according to the applied sequence of unitary operations, a measurement is performed on the resulting state which is mapped to create some artefact.  Like their classical counterparts, the expressive power and utility of quantum grammars are contingent on the careful design of rules and symbol mappings. The DisCoCat (Distributional Compositional Categorial) framework offers an example of a quantum-inspired grammatical approach that has been applied to natural language processing and generation \cite{karamlouQuantumNaturalLanguage2022}, illustrating the potential for such formalisms.

\textbf{Automata} are systems that transition between states based on a set of rules applied to their current state and, potentially, input. They share conceptual similarities with grammars in their rule-based, step-by-step evolution.

\subsubsection{Classical Automata}
The most widely known examples of these are \emph{Cellular Automata} (CA), with Conway's Game of Life being a canonical instance. In CAs, a grid of cells, each in a particular state, evolves synchronously based on rules that consider the states of neighboring cells. While CAs are often designed with local interactions (cells only interacting with their immediate neighbors), there are no fundamental restrictions preventing non-local interactions. The design of the rules determines the complexity and behavior of the automaton, allowing them to be computationally efficient or arbitrarily complex. In PCG, CAs are foundational and have been employed for diverse tasks, including terrain generation, the creation of dungeon-like structures (mazes, caves), and the placement and generation of various game assets \cite{shakerProceduralContentGeneration2016}.

\subsubsection{Quantum Automata} These are another approach to automata \cite{mooreQuantumAutomataQuantum2000} which is more of an analogy than a replacement to its classical counterpart. In prototypical quantum cellular automata, an array of quantum subsystems (each typically comprising one or more qubits) evolves under the repeated application of a global unitary operation $U$, which propagates interactions between neighboring subsystems. The definition of `neighborhood' is flexible, similar to classical CAs, and dictates the pattern of entanglement and information flow. At any iteration, the state of these subsystems can be measured, yielding classical data that can then be interpreted or translated to generate the desired PCG artefact. A key distinction is that a Quantum CA will evolve into a superposition of states. This opens up the possibility of employing QAs to selectively amplify basis states that satisfy certain desirable criteria before measurement. While this amplification step could offer a quantum advantage for specific objectives, it is likely that the patterns and structures generated by QCAs will be inherently different from those produced by classical automata, offering a rich, largely unexplored design space for novel PCG outcomes.

\section{Specific examples}
\subsection{Likely integrations of quantum algorithms in PCG}

\begin{figure}
    \centering
    \includegraphics[width=0.5\textwidth]{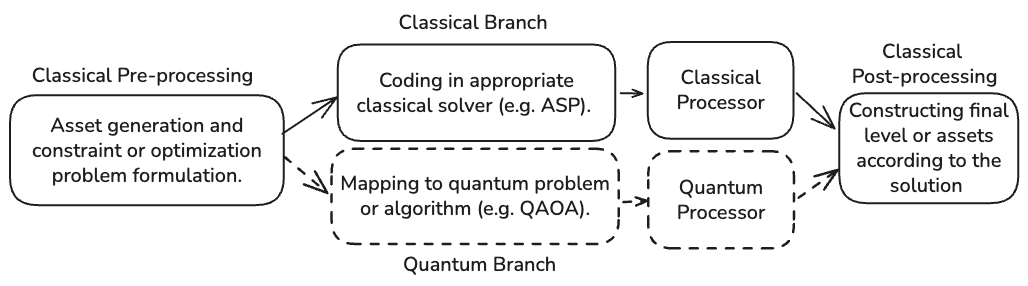}
    \caption{The first implementations of quantum computers will likely just offload a computationally complex task like the solver for a given problem.} 
    \label{fig:replacement}
\end{figure}

The most likely initial implementations of QCs in any PCG task will be to replace a computationally expensive subroutine with a QA that has a known quantum advantage, much like GPUs. As such, the routines used to generate a world or level simply would call a QC rather than a CPU to compute much more complicated conditions. Figure \ref{fig:replacement} illustrates this. This is a very generic and simple example, but specific implementations would require dedicated research. 

Indeed, this is already possible (although with lower quality solutions than using a CC), and has been shown in the context of WFC \cite{merrellModelSynthesisGeneral2011}. Here, the solver portion of the algorithm has been replaced by a probabilistic system in the case of Heese \cite{heeseQuantumWaveFunction2024} or a QAOA in the case of Zhang \cite{zhangWaveFunctionCollapse2024}. Although neither of these techniques have an advantage, they do show that in principle such an approach can work. Similarly, in the case of example based procedural modeling \cite{merrellExampleBasedProceduralModeling2023}, one could discover a QA that solves the problem of computing the graph hierearchy, replacing the current classical approach. However, all of this is currently speculative, however, it is possible to think of a game that requires existign QAs. 

\subsection{Gedankenspiel}

In theoretical physics it is common to consider a Gedankenexperiment, so in theoretical game design it is only fair to provide some ideas on proof-of-principle games that could only exist with a fault-tolerant quantum advantage. Though real-life quantum-powered games will likely seek to create engaging content, this Gedankenspiel focuses on illustrating a concrete quantum advantage over prioritizing fun. This requires live queries to a QC, but the algorithm is very efficient, so this may even be possible on an early-FTQC for classically complex knots. 

\emph{Knotty Jones} exploits a QA which computes the expensive Jones polynomial (JP) of a knot \cite{laakkonenLessQuantumMore2025}. The JP is an invariant of any given knot, although it is not necessarily the case that two different knots will have different JPs. 

\begin{figure}
        \centering
        \includegraphics[width=0.4\textwidth]{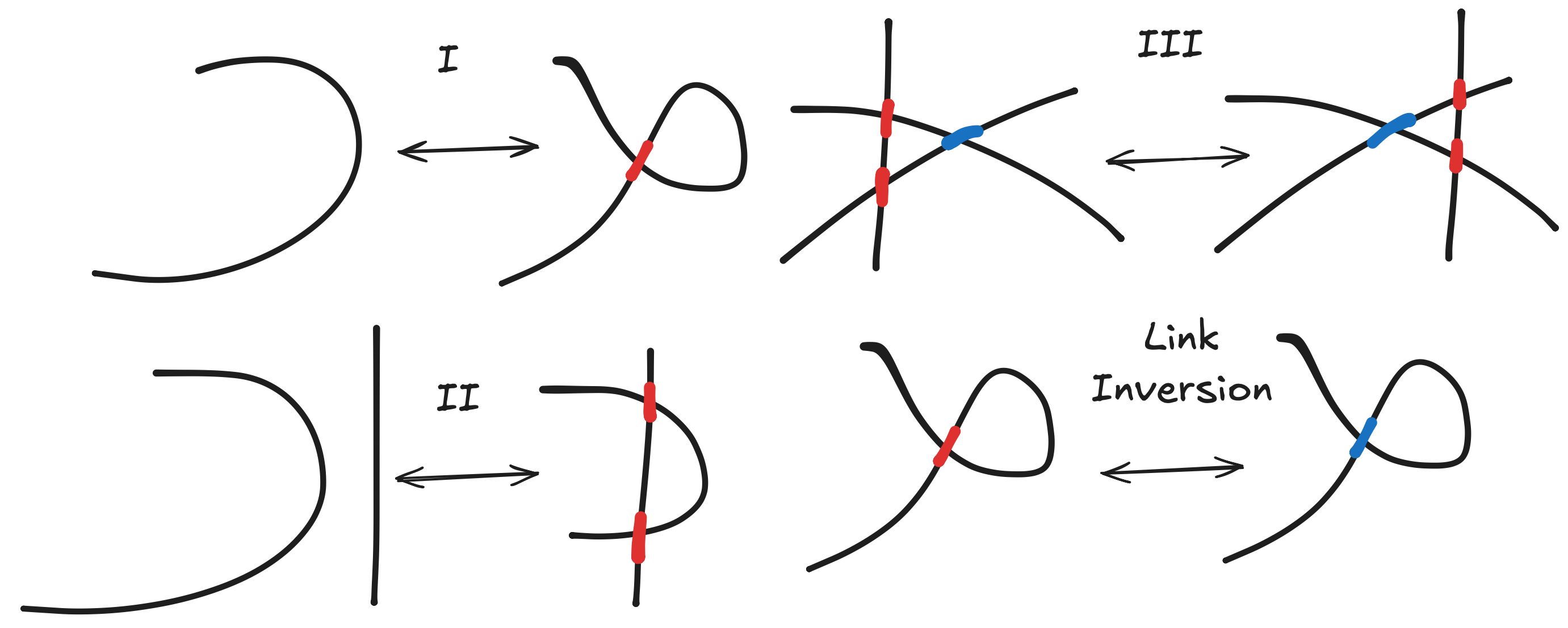}
        \caption{The three types of Reidermeister moves. Coloured crossings represent whether the knot is over (red) or under (blue) the crossing, with the strand in question being the one with the most overlap with the colouring. Reidermeister moves do not change invariants of the knot, that is to say, they cannot untangle a knot beyond a certain number of a knot's inherent irreducible crossings.} 
        \label{fig:knot}
 \end{figure}

\begin{figure}[b]
    \centering
    \includegraphics[width=0.4\textwidth]{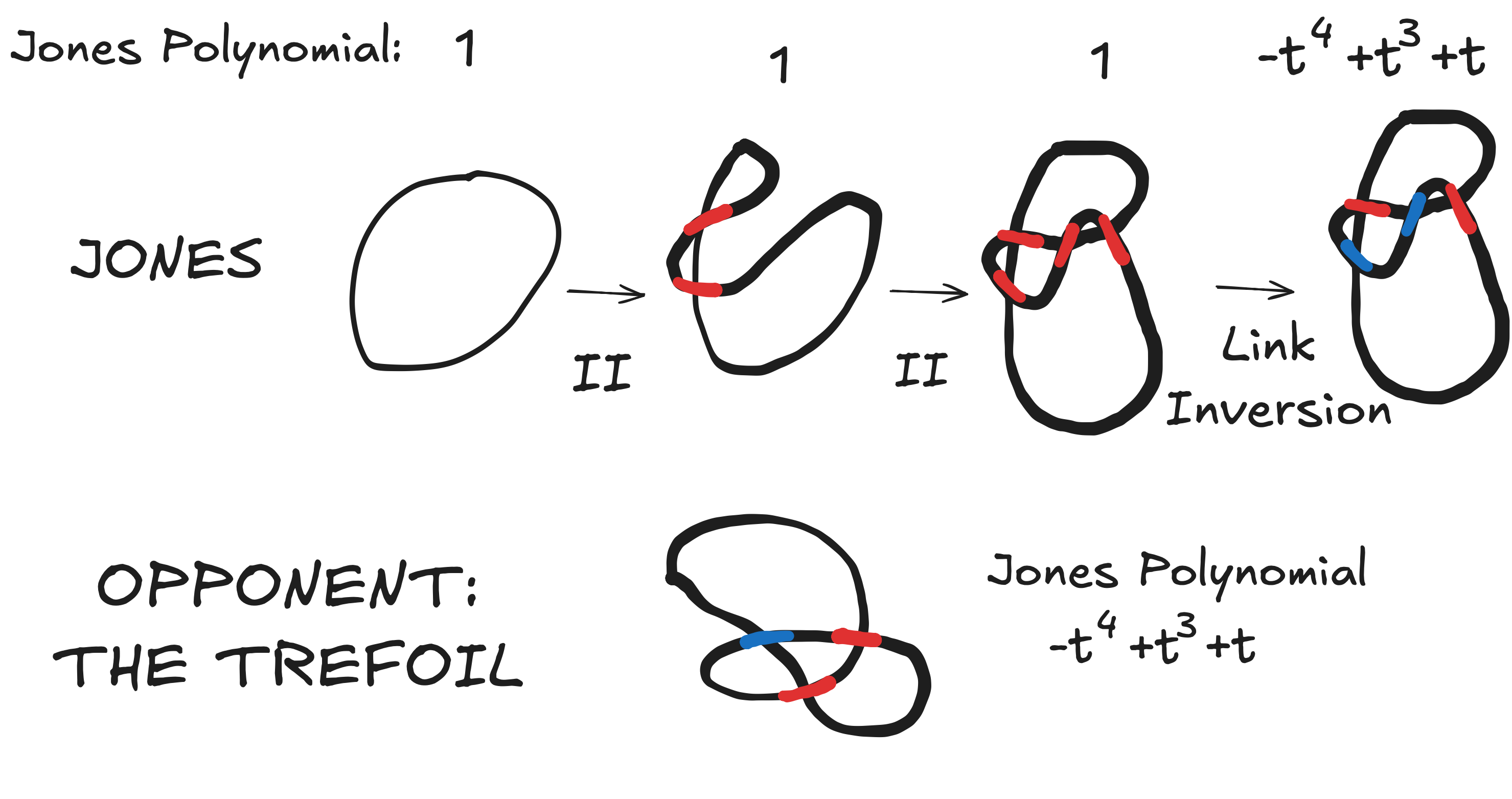}
    \caption{A turn of \textit{Knotty Jones} where the player is fighting a simple opponent in the form of the trefoil knot. Jones plays a sequence of two Reidermeister moves and two link inversions to transform itself into an equivalent knot to the opponent's, thus quickly winning the round as equivalent knots have the same Jones polynomial. The knots shown here are very simple, computational advantage would only occur for knots with many more crossings.} 
    \label{fig:knottyjones}
\end{figure}

Knotty Jones has you playing as a young knot called Jones that begins as the simplest knot (the unknot with JP=1) and battles other knots of increasing complexity. A turn consists of performing a finite number of Reidermeister moves (these do not change the JP) and some link inversions as in Figure \ref{fig:knot}, which do change the JP.  The QC then computes the JP. The player wins if their moves transform their knot to have the same JP as the opponent. An example of a turn is given in Figure \ref{fig:knottyjones}. 

 Although at lower knot complexities,\footnote{Unrigorously defined as the number of crossings that cannot be reduced by Reidermeister moves.} most knots with the same JP are equivalent: there is  a sequence of Reidermeister moves that transform between two equivalent knots. As the complexity of the knots increases, there are examples of knots that have the same JP but are not equivalent, such as the Conway knot and the Kinoshita–Terasaka knot. Here is where the game becomes more non-trivial and the advantage of a QA becomes essential for the game. This game conveniently circumvents the limitations of using the JP as an invariant for two reasons: 1) One does not need to ensure that the knots are equivalent. 2) The generation of opponent knots must only be post-selected to ensure that the JP is different from the player's.

 Thus, the PCG of knots can be done classically by making a sequence of random Reidermeister moves interlaced with link inversions, and the QC can then be used to compute the JP of the knot. Some post-selection criteria could be the number of terms in the Jones polynomial and, of course, that it be different from the current JP of the player. This would be extremely inefficient classically as computing the JP is \#P-hard, which is a different class of problems from NP-complete, but are still classically inefficient \cite{laakkonenLessQuantumMore2025}. For scale, an A100 is memory limited to $\approx1500$ crossings in $\approx5$ minutes. The QA running on a superconducting QC takes $\approx1$ minute, but this remains the similar for much larger knots due to the QA's logarithmic scaling. 

The generalization of this game is any game where one wants to score the player based on the closeness of their solution to the most optimal solution. If the puzzle in question has no efficient classically computable solution, then it is possible to use a QC to take this role. For problems like the travelling salesman, where it is not thought that an efficient quantum solution exists, it still may be possible to compute better solutions to compare against than classical methods allow. For PCG specifically, the same idea can be ported to generate-and-reject type processes where the fitness function can be made more complex. 

\section{Conclusion}

Quantum computing is still an emergent technology, and the noisy machines that currently exist are not appropriate for many tasks in PCG that require exactness. Fortunately, it is possible to create error-corrected QCs, and this is beginning to become experimentally possible, which promises to bring in the era of (early) fault tolerance. This paper has described what an advantage is in this era, the practical nuances associated with such a term, and provided several examples of where quantum algorithms could be used in PCG, both from the perspective of advantage and novelty. Finally, a Gedankenspiel with a simple PCG example that would be impossible at higher complexities without a significant quantum advantage was proposed alongside some other potential avenues of exploration.  

Though fault-tolerant quantum computing has very promising avenues of exploration overall, it does not by itself overcome fundamental challenges in PCG that are more creative in nature than algorithmic. Nevertheless, by exploiting the significant advantages we currently know, and finding specific new advantages for specific problem types, we can hope to bring about a plethora of exciting new approaches to PCG. 
\printbibliography

\end{document}